\documentclass[a4paper,11pt]{article}
\pdfoutput=1

\usepackage{jcappub}
\usepackage[T1]{fontenc}

\title{Cosmographic connection between cosmological and Planck scales: the Barrow--Tsallis entropy}

\author[a]{Yu. L. Bolotin,}
\author[b,c]{V. V. Yanovsky,}
\author[b]{and D. A. Yerokhin}

\affiliation[a]{National Science Center Kharkov Institute of Physics and Technology,\\
1 Akademicheskaya str., Kharkov, 61108, Ukraine}
\affiliation[b]{V. N. Karazin Kharkiv National University,\\
4 Svobody Sq., Kharkiv, 61022, Ukraine}
\affiliation[c]{Institute for Single Crystals, NAS Ukraine,\\
60 Nauky Ave., Kharkov, 31001, Ukraine}

\emailAdd{ybolotin@gmail.com}
\emailAdd{yanovsky@isc.kharkov.ua}
\emailAdd{denyerokhin@gmail.com}

\abstract{One of the fundamental challenges of quantum gravity is to understand how the microscopic degrees of freedom of the cosmological horizon shape the evolution 
of the Universe. One possible approach to this problem is based on the Barrow--Tsallis entropy. This entropy accounts for both quantum gravitational 
effects and the nonextensive effects inherent in any long-range interaction. By employing an inverse cosmographic reconstruction of the model parameters, we derive a relation between the Barrow parameter, which encodes the microscopic deformation of the horizon geometry, and the Tsallis parameter, which characterizes macroscopic nonextensivity. Within the IR--UV correspondence, this relation determines the scaling of the microscopic length uncertainty in terms of the current cosmographic parameters and demonstrates how long-range nonextensive effects alter the standard Karolyhazy-type scaling. We also applied our cosmographic reconstruction method to evaluate the 
feasibility of using fractional derivatives to describe the late evolution of the Universe. Within the assumed non-interacting power-law holographic model class, the resulting algebraic relations are exact. For this fixed model class, the observational uncertainty of the reconstructed parameter combination is determined by the current uncertainties in the cosmographic parameters; the quoted uncertainty of $\delta$ additionally includes the adopted prior on $\Delta$, but not uncertainty associated with the model choice. Propagating the observational errors of the deceleration and jerk parameters and marginalizing over a uniform prior on the Barrow parameter within the adopted interval $\Delta \in [-1,1]$, we obtain the Monte Carlo estimate $\delta = 1.11 \pm 0.57$ for the nonextensivity parameter, with the jerk parameter providing the largest observational contribution to the error budget.}


\begin{document}
\maketitle
\flushbottom

\section{Barrow--Tsallis entropy and cosmographic constraints}
The conceptual and  observational problems of the Standard  Cosmological Model (SCM) \cite{Bull:2015stt,Peebles:2002gy}  and the long-standing futile attempts to directly capture the model's main components (dark energy and dark matter) \cite{Copeland:2006wr,Frieman:2008sn}  have led to the desire, at least at the phenomenological level, to construct a cosmological model representing a synthesis of gravity and thermodynamics and not including components of unknown nature. It is precisely this model -- entropic cosmology -- that has achieved real success in describing the evolution of the Universe \cite{Easson:2010av,Easson:2010xf,Padmanabhan:2003gd,Padmanabhan:2009vy,Nojiri:2024zdu,Bolotin:2023wiw}. 
As physics evolved, it came to consider systems that differed dramatically in both their spatial and temporal scales and the nature of interactions between individual parts of the system. This diversity precludes the use of a single definition of entropy at the phenomenological level and has led to the emergence of a number of alternative formulations of this fundamental concept \cite{Bekenstein:1973ur,Hawking:1976de,Tsallis:1987eu,Barrow:2020tzx,Nojiri:2022aof}.

Each new generalization of the standard entropy of the black hole event horizon (Bekenstein entropy) aims to take into account specific physical effects. Thus, the Barrow entropy \cite{Barrow:2020tzx} (we use a system of units in which  $8\pi G=c=k_B=1$)
\begin{equation}
	S_B=\left(\frac{A}{4A_P}\right)^{1+\frac{\Delta}{2}}
\end{equation}
was introduced to account for quantum gravitational effects leading to deformation of the black hole's event horizon (here $\Delta$ is the deformation parameter). In the limit $\Delta =0$ the standard Bekenstein--Hawking entropy is restored, while $\Delta =1$ corresponds to the maximum deformation.

Although Barrow originally justified the parameter $\Delta$ through a simple fractal model --- namely, the spherical analogue of a Koch snowflake construction --- which naturally limits $\Delta$ to the interval $[0,1]$, this range can be extended to negative values in a purely phenomenological sense: surface geometries characterized by voids or internal porosity, such as spongy or porous structures, can exhibit effective fractal dimensions smaller than their Euclidean dimension.

Since $S_B \propto L^{2+\Delta}$, the exponent defines an effective horizon dimension $D_{eff}=2+\Delta$. Requiring the deformed horizon to remain between a line and a volume, $1 \le D_{eff} \le 3$, gives the interval $-1 \le \Delta \le 1$ adopted in this work. Thus, $0 \le \Delta \le 1$ is the original Barrow sector, whereas $-1 \le \Delta <0$ is the phenomenological porous-horizon extension.

The Tsallis entropy \cite{Tsallis:1987eu} is a non-additive generalization of the 
Boltzmann--Gibbs entropy for systems with long-range interactions. The Tsallis entropy is related to the Bekenstein entropy by the relation 
\begin{equation}\label{S_T}
	S_T = (S_{BH})^\delta
\end{equation}
where $\delta \in (0,\infty)$ characterizes the degree of nonextensivity. This restriction ensures that $S_T$ increases monotonically with $S_{BH}$; $\delta=0$ would give a constant entropy, whereas $\delta<0$ would make the entropy decrease with the horizon area. In the case $\delta=1$, the standard Bekenstein--Hawking entropy is restored. Beyond the positivity condition, the entropy definition alone imposes no finite upper bound on $\delta$; the value $\delta = 3/2$ (at $\Delta = 0$) corresponds to the extensive limit of the horizon entropy \cite{Jizba:2023fkp} considered below. A finite range of $\delta$ arises only a posteriori: the cosmographic constraint \eqref{alpha_q0j0Delta}, combined with the interval $-1 \le \Delta \le 1$ adopted above, selects $0.66 \lesssim \delta \lesssim 1.97$ for the central values of the reference cosmographic parameters, while the wider intervals of $\delta$ shown in the figures are for illustration only.

The main goal of quantum gravity is to understand how microscopic 
gravitational degrees of freedom can shape macroscopic evolution. 
Some light on this question can be shed by the Barrow--Tsallis entropy, 
which takes into account the deformation of the horizon due to both 
quantum and long-range gravitational interactions. Such a generalized 
entropy can be constructed in the following way \cite{Yarahmadi:2025ema}. First, we deform 
the area of the horizon due to quantum effects $A \to A^{1 + \frac{\Delta}{2}}$ and then use this expression as the area included in the Tsallis entropy \eqref{S_T}. 
The generalized Barrow--Tsallis entropy $S_{BT}$ constructed in this way 
has the form
\begin{equation}\label{S_BT}
	S_{BT} = \gamma \left( \frac{A}{A_P} \right)^{\left( 1 + \frac{\Delta}{2} \right) \delta}
\end{equation}

Here $\gamma>0$ is a dimensionless normalization constant. It is not fixed uniquely by the construction: the deformation $A \to A^{1 + \frac{\Delta}{2}}$ and the subsequent Tsallis composition determine only the exponent of the entropy, while the overall normalization must be inherited from the underlying microscopic theory. If eq.~\eqref{S_BT} is obtained literally by raising the Barrow entropy defined above to the power $\delta$, the corresponding choice is $\gamma=4^{-(1+\Delta/2)\delta}$; in particular, the undeformed Bekenstein--Hawking limit gives $\gamma=1/4$. Since the analysis below concerns only the scaling with $A$, and the horizon-entropy arguments of section~3 are formulated at the level of order-of-magnitude inequalities, all numerical geometric factors are absorbed into $\gamma$, and we use the convention $\gamma=1$. This constant changes only the overall normalization of the entropy and the $\mathcal{O}(1)$ prefactor $\gamma^{-1/3}$ of the microscopic length bound \eqref{delta_L_BT}; it does not change the scaling exponents or the cosmographic relation between $\Delta$ and $\delta$.

\begin{figure}[tbp]
	\centering
	\includegraphics[width=0.8\textwidth]{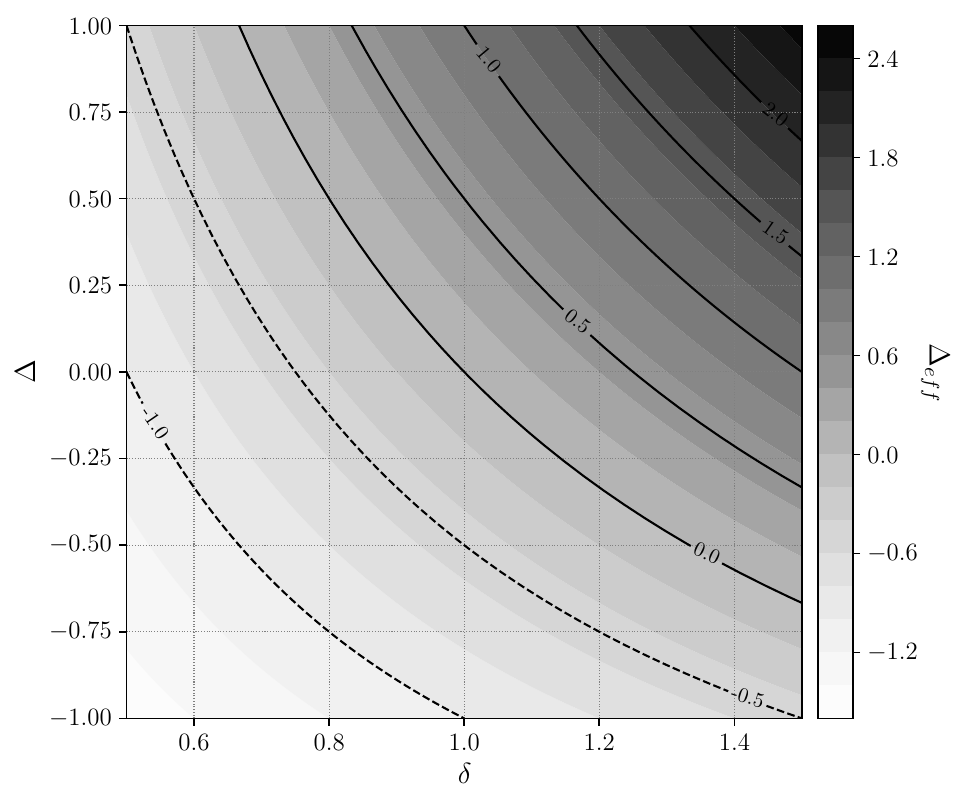}
	\caption{Dependence of the effective parameter $\Delta_{eff}$ on the parameters $\Delta$ and $\delta$.}
	\label{fig:delta_eff}
\end{figure}

Note that the Barrow--Tsallis entropy can be given the form of the Barrow entropy by introducing the effective Barrow exponent $\Delta_{eff} = 2(\delta - 1) + \Delta \delta.$ Indeed, for the Barrow--Tsallis entropy we obtain
\begin{equation}
	S_{BT} = \gamma \left( \frac{A}{A_P} \right)^{1 + \frac{\Delta_{eff}}{2}}
\end{equation}
The difference lies in the range of changes in the effective parameter 
$\Delta_{eff}$, which is determined by changes in $\Delta$ and $\delta$. 
The dependence of this parameter on $\Delta$ and $\delta$ is shown in 
figure~\ref{fig:delta_eff}.

The Barrow--Tsallis entropy \eqref{S_BT} allows one, within the framework of entropic cosmology, to construct in a standard way holographic dark energy \cite{Wang:2016och,Saridakis:2020lrg,Saridakis:2020zol} and modified Friedmann equations that take into account both corrections to cosmological dynamics generated by quantum fluctuations and nonextensivity associated with the long-range nature of gravity.

The density of holographic dark energy generated by the Barrow--Tsallis 
entropy can be represented in the form \footnote{It is worth noting that $\alpha$ is simply related to the effective Barrow exponent as $\alpha = 2 - \Delta_{eff}$. This highlights that the energy density scaling is entirely determined by the horizon's departure from the standard Bekenstein--Hawking area law.}
\begin{equation}
	\label{rho_de_alpha}
	\rho_{de} = 3\beta H^\alpha, \quad \alpha = 4 - 2\delta - \delta\Delta
\end{equation}

For a universe filled with holographic dark energy with density $\rho_{de}$, 
pressure $p_{de}$ and cold dark matter with density $\rho_m$ (and zero pressure), the Friedmann equations take the form
\begin{equation}	\label{Frid_I}
	H^2 = \frac{1}{3} (\rho_m + \rho_{de})
\end{equation}
\begin{equation}\label{Frid_II}
	\dot{H} = -\frac{1}{2} (\rho_m + \rho_{de} + p_{de})
\end{equation}

The conservation equations for the case of non-interacting energy components read

\begin{equation}\label{Cons_rho_m}
	\dot{\rho}_m + 3H\rho_m = 0
\end{equation}
\begin{equation}\label{Cons_rho_de}
	\dot{\rho}_{de} + 3H(\rho_{de} + p_{de}) = 0
\end{equation}

We emphasize that the algebraic exactness used below is conditional on the power-law ansatz \eqref{rho_de_alpha}, spatial flatness, constant $\alpha$ and $\beta$, and the separate conservation laws \eqref{Cons_rho_m}--\eqref{Cons_rho_de}; it does not imply model independence with respect to arbitrary dark-sector dynamics.

Using \eqref{rho_de_alpha},\eqref{Frid_I} and \eqref{Cons_rho_de}, we find

\begin{equation}\label{rho_m}
	\rho_m = 3H^2 - 3\beta H^\alpha
\end{equation}
\begin{equation}\label{p_de}
	p_{de} = -3\beta H^\alpha - \alpha \beta H^{\alpha - 2} \dot{H}
\end{equation}

Differentiating equation \eqref{Frid_II} with respect to time, we obtain a closed system 
of equations for determining the model parameters $\alpha$ and $\beta$:
\begin{equation}\label{alpha_beta}
	\left( 1 - \frac{1}{2} \alpha \beta H^{\alpha-2} \right) \dot{H} = -\frac{3}{2} (H^2 - \beta H^\alpha)
\end{equation}
\begin{equation}\label{beta_alpha}
	\left( 1 - \frac{1}{2} \alpha \beta H^{\alpha-2} \right) \ddot{H} - \frac{1}{2} \alpha \beta (\alpha - 2) H^{\alpha-3} \dot{H}^2 = -\frac{3}{2} \dot{H} (2H - \alpha \beta H^{\alpha-1})
\end{equation}

Equation \eqref{beta_alpha} is the time derivative of \eqref{alpha_beta}, serving as a second local kinematic condition. We formulate an inverse cosmographic problem: for a given local set $(H, \dot{H}, \ddot{H})$, the time-independent model parameters $\alpha$ and $\beta$ are reconstructed algebraically. Throughout the paper, the term ``exact'' refers precisely to this algebraic inversion within the adopted model class; the IR--UV correspondence invoked in section~3 constitutes an additional physical hypothesis. The resulting expressions depend on the dimensionless kinematic combinations $\dot{H}/H^2$ and $\ddot{H}/H^3$:
\begin{equation}\label{alpha}
	\alpha = -\frac{9 \frac{\dot{H}}{H^2} + 3 \frac{\ddot{H}}{H^3}}{\left( \frac{\dot{H}}{H^2} \right)^2 \left( 2 \frac{\dot{H}}{H^2} + 3 \right)}
\end{equation}
\begin{equation}\label{beta}
	\beta = H^{2-\alpha} \frac{2 \frac{\dot{H}}{H^2} + 3}{\alpha \frac{\dot{H}}{H^2} + 3}
\end{equation}

The dimensionless combinations $\dot{H}/H^2$ and $\ddot{H}/H^3$ are directly related to the cosmographic parameters \cite{Bolotin:2018xtq}:
\begin{equation}\label{dot_H_q}
	\frac{\dot{H}}{H^2} = -(1 + q)
\end{equation}
\begin{equation}\label{ddot_H_q_j}
	\frac{\ddot{H}}{H^3} = j + 3q + 2
\end{equation}
where $q = -\frac{\ddot{a}}{aH^2}$ is the deceleration parameter, and $j = \frac{\dddot{a}}{aH^3}$ is the jerk parameter.

Note that from \eqref{rho_de_alpha} and \eqref{p_de} follows the expression for the parameter of the equation of state of holographic dark energy:

\begin{equation}
	w_{de} \equiv \frac{p_{de}}{\rho_{de}} = -1 - \frac{\alpha}{3} \frac{\dot{H}}{H^2}
\end{equation}
Taking into account relation \eqref{dot_H_q}, this parameter can be rewritten in the form:
\begin{equation}
	w_{de} = -1 + \frac{\alpha}{3} (1 + q)
\end{equation}

Substituting \eqref{dot_H_q} and \eqref{ddot_H_q_j} into \eqref{alpha} and \eqref{beta} we find
\begin{equation}\label{alpha_qj}
	\alpha(q, j) = \frac{3}{2} \frac{j-1}{(q+1)^2 (q - 1/2)}
\end{equation}
\begin{equation}\label{beta_qj}
	\beta(H, q, j) = H^{2-\alpha(q, j)} \frac{1-2q}{3 - \alpha(q, j)(1+q)}
\end{equation}

The model parameters $\alpha$ and $\beta$ are constants, and the 
cosmographic parameters, generally speaking, are functions of time. 
However, it can be shown \cite{Bolotin:2018xtq} that for expressions \eqref{alpha_qj}, \eqref{beta_qj}
$\frac{d\alpha}{dt}=0$ and $\frac{d\beta}{dt}=0$. Consequently, the right-hand sides in \eqref{alpha_qj} and \eqref{beta_qj} can be evaluated at any epoch where the corresponding cosmographic data are available. In the present work, we calculate them at $z=0$. This approach determines the model parameters strictly through the local cosmographic data $(H_0, q_0, j_0)$, independent of any global ansatz for the expansion history. Using the current values $q_0$ and $j_0$, we write \eqref{alpha_qj} as
\begin{equation}\label{alpha_q0j0}
	\alpha(q_0, j_0) = \frac{3}{2} \frac{j_0 - 1}{(q_0 + 1)^2 \left( q_0 - \frac{1}{2} \right)}
\end{equation}

To examine the validity of the procedure used, let us move on to 
dimensionless density parameters $\Omega_m = \rho_m / (3H^2)$ and 
$\Omega_{de} = 1 - \Omega_m$. Using the modified Friedmann equation \eqref{Frid_I}
together with the conservation equations \eqref{Cons_rho_m}--\eqref{Cons_rho_de}, the deceleration 
parameter can be written as

\begin{equation}
	q(\Omega_m; \alpha) = -1 + \frac{3}{2} \frac{\Omega_m}{1 - \frac{\alpha}{2}(1 - \Omega_m)}
\end{equation}
This relation allows us to analyze the model's behavior at the limiting points of evolution and examine potential singularities in eq.~\eqref{alpha_qj}.

The expression for $\alpha(q, j)$ contains formal singularities corresponding to the limiting regimes of the Universe's evolution. Consider the de Sitter branch $q \to -1$ ($\Omega_m \to 0$) of the non-interacting dust plus holographic-dark-energy model with constant finite $\alpha\neq2$ \cite{Bolotin:2018xtq,Bolotin:2015dja,Chakraborty:2025rvc}. Differentiating \eqref{dot_H_q} with respect to $N \equiv \ln a$ and using \eqref{ddot_H_q_j}, one obtains the exact kinematic identity
\begin{equation}\label{jerk_identity}
	j - 1 = 2(1+q)^2 - 3(1+q) - \frac{d(1+q)}{dN},
\end{equation}
valid for any expansion history. Near the de Sitter state the Hubble parameter tends to a constant, while the matter density dilutes as $\rho_m \propto a^{-3}$ by virtue of \eqref{Cons_rho_m}; expanding $q(\Omega_m; \alpha)$ around $\Omega_m = 0$ gives $1+q = \frac{3}{2-\alpha}\,\Omega_m + \mathcal{O}(\Omega_m^2)$, and hence $d(1+q)/dN = -3(1+q) + \mathcal{O}\big((1+q)^2\big)$. The linear terms in \eqref{jerk_identity} cancel identically, leaving $j-1=\mathcal{O}\big((1+q)^2\big)$; carrying the expansion to second order fixes the coefficient,
\begin{equation*}
	j-1 = -\frac{9\alpha}{(2-\alpha)^2}\,\Omega_m^2+\mathcal{O}(\Omega_m^3) = -\alpha(1+q)^2+\mathcal{O}\big((1+q)^3\big).
\end{equation*}
Thus, the numerator of eq.~\eqref{alpha_qj} has a second-order zero that exactly compensates the second-order zero of the denominator, and $\lim\limits_{q \to -1} \alpha(q, j)=\alpha$. We stress that the quadratic (rather than linear) behavior relies on the pressureless character of the diluting matter component: for a component with an equation of state $w_m \neq 0$ one would instead have $1+q \propto a^{-3(1+w_m)}$ and, consequently, $j - 1 = 3 w_m (1+q) + \mathcal{O}\big((1+q)^2\big)$. The quadratic law thus follows from the assumed field equations and separate conservation laws; it is not a general cosmographic identity.

A second singularity arises at $q = 1/2$, corresponding to the matter-dominated epoch ($\Omega_{de} \to 0$). In the vicinity of this solution, an expansion of the cosmographic parameters in terms of the small dark energy density leads to a linear relation $j-1 \propto (q - 1/2)$. Substituting this expansion into \eqref{alpha_qj} removes the apparent singularity of the function $\alpha(q, j)$, rendering it an analytic function throughout the entire evolutionary history of the Universe.

Relation \eqref{alpha_q0j0}, together with \eqref{rho_de_alpha}, imposes a constraint on the entropy parameters. By fixing the combination $\delta(2+\Delta)$, this condition reduces the two-parameter Barrow--Tsallis entropic dark energy model $[\Delta,\delta]$ to a one-parameter family $[\Delta,\delta(\Delta)]$.

Let us demonstrate this relationship using the special case of the extensive 
limit of the Barrow--Tsallis entropy \cite{Jizba:2023fkp}. The requirement of 
extensiveness of the entropy \eqref{S_BT} leads to a simple relationship between 
the parameters $\Delta$ and $\delta$:
\begin{equation}
	\left( 1 + \frac{\Delta}{2} \right) \delta = \frac{3}{2} \implies \delta = \frac{3}{2 + \Delta}
\end{equation}
The expression \eqref{alpha_q0j0} obtained for $\alpha(q_0, j_0)$ allows us to establish 
this relationship in the general case. Using \eqref{rho_de_alpha}, we find the 
relationship we are interested in between parameters $\Delta$ and $\delta$:
\begin{equation}\label{alpha_q0j0Delta}
	4 - 2\delta - \delta \Delta = \alpha(q_0, j_0)
\end{equation}
Solving \eqref{alpha_q0j0Delta} for $\Delta$, we obtain:
\begin{equation}
	\Delta(\delta; q_0, j_0) = \frac{4 - \alpha(q_0, j_0)}{\delta} - 2 = \frac{4 - 2\delta}{\delta} + \frac{3(j_0 - 1)}{(1 + q_0)^2 (1 - 2q_0)\delta}
\end{equation}
Equivalently, $\delta(\Delta;q_0,j_0)=[4-\alpha(q_0,j_0)]/(2+\Delta)$. Since $2+\Delta\in[1,3]$, the physical branch $\delta>0$ requires $\alpha(q_0,j_0)<4$; for fixed cosmographic parameters its allowed interval is $[4-\alpha(q_0,j_0)]/3\le\delta\le4-\alpha(q_0,j_0)$.

Within the model defined by eqs.~\eqref{rho_de_alpha}--\eqref{Cons_rho_de}, relation \eqref{alpha_q0j0} follows algebraically without approximation. With this model class fixed, the observational uncertainty of the reconstructed combination is driven by the current errors in the cosmographic parameters $q_0$ and $j_0$. Currently, the deceleration parameter $q_0$ is fairly well constrained \cite{Mukherjee:2020vkx}; the non-parametric Gaussian-process reconstruction of $q(z)$ in ref.~\cite{Mukherjee:2020vkx} yields, depending on the combination of datasets used (cosmic chronometers with two alternative stellar-population-synthesis calibrations, combined with the Pantheon supernovae, without and with BAO data --- the sets N1--N4 of that reference, for a spatially flat universe):

\begin{equation}\label{q0_values}
	q_0 = -0.573^{+0.041}_{-0.042}, \quad -0.580^{+0.055}_{-0.063}, \quad -0.533^{+0.038}_{-0.038}, \quad -0.574^{+0.044}_{-0.045}
\end{equation}

This result is in good agreement with the current value of the deceleration 
parameter in SCM:
\begin{equation}
	q_0 = \frac{1 - 3\Omega_{\Lambda 0}}{2} \approx -0.6
\end{equation}
The situation with determining the current value of the jerk parameter is much worse \cite{AlMamon:2018uby,Zhai:2013fxa}: the spread of $j_0$ between different parameterizations and datasets reaches approximately 100\%. For the jerk parameter we adopt the observational determination
\begin{equation}\label{j0_observational}
    j_0=0.742^{+0.106}_{-0.094},
\end{equation}
reported in table~I of ref.~\cite{AlMamon:2018uby} and obtained by fitting the logarithmic parameterization of the deceleration parameter, $q(z)=q_0+q_1\ln(1+z)$, to 41 measurements of the Hubble parameter in the range $0.07 \le z \le 2.36$. This choice is internally consistent with the reconstructions behind eq.~\eqref{q0_values}: the deceleration parameter of the same fit, $q_0=-0.466^{+0.093}_{-0.096}$, agrees within $1\sigma$ with all four values in eq.~\eqref{q0_values}. We stress that the quoted error of $j_0$ is the statistical uncertainty conditional on the adopted parameterization, while the spread of the central $j_0$ values across the parameterization family of ref.~\cite{AlMamon:2018uby} constitutes the $\sim 100\%$ model dependence mentioned above. However, 
improving the accuracy of determining the parameter is only a matter of time. 
In particular, cosmography using next-generation gravitational wave 
detectors \cite{Chen:2024gdn}  will significantly improve the accuracy of determining 
cosmographic parameters. Recall that the first measurements of the Hubble 
parameter differed by an order of magnitude from modern ones. Achieving the 
required accuracy will solve the problem of finding the parameters of 
two-parameter cosmological models. As the reference pair for the numerical illustrations below we use $q_0=-0.580^{+0.055}_{-0.063}$ --- the most conservative of the four determinations~\eqref{q0_values} in terms of the quoted uncertainty --- together with the jerk value~\eqref{j0_observational}; the robustness of the results with respect to this choice is quantified in table~\ref{tab:MC_delta}. 

Models with a larger number of parameters will require higher-order cosmographic parameters. In conditions where cosmography fixes only a combination of parameters $\delta(2 + \Delta)$, estimation of the nonextensivity parameter $\delta$ is possible only with an a priori assignment of the distribution of the Barrow parameter $\Delta$.

\begin{figure}[tbp]
	\centering
	\includegraphics[width=0.8\textwidth]{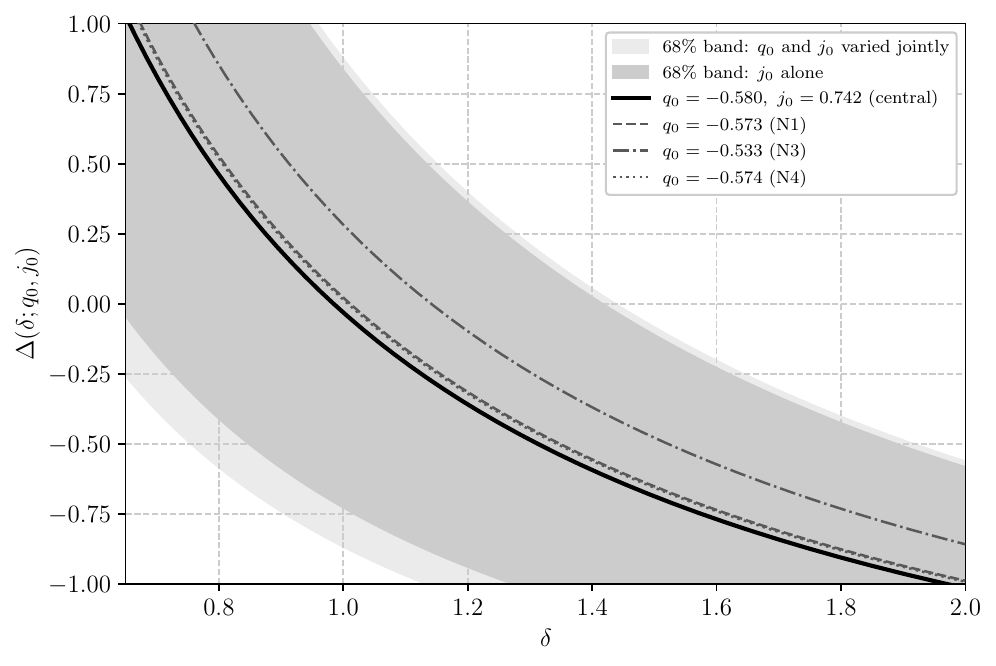}
	\caption{Barrow parameter $\Delta$ as a function of the nonextensivity parameter $\delta$. The solid curve corresponds to the central values of the reference cosmographic parameters, $q_0 = -0.580$ \cite{Mukherjee:2020vkx} and $j_0 = 0.742$ \cite{AlMamon:2018uby}. The outer (light) band is the 68\% region obtained by propagating the observational uncertainties of $q_0$ and $j_0$ jointly [eqs.~\eqref{q0_values} and~\eqref{j0_observational}], with the same conventions as in the Monte Carlo analysis below; the inner (darker) band is the 68\% region produced by the jerk uncertainty alone, showing that $j_0$ dominates the width of the allowed region. The thin curves correspond to the central values of the three alternative determinations of $q_0$ in eq.~\eqref{q0_values}; their own jerk bands are of nearly the same width and are omitted for clarity.}
	\label{fig:Delta_delta}
\end{figure}

For each fixed pair $(q_0, j_0)$, the corresponding curve $\delta(\Delta; q_0, j_0)$ is defined by eq.~\eqref{alpha_q0j0Delta}. For the Barrow parameter we adopt a uniform prior over the extended interval $\Delta \in [-1,1]$ motivated above by $1 \le D_{eff} \le 3$; this includes the original Barrow sector $[0,1]$ and the phenomenological porous-horizon extension $[-1,0)$, and no additional narrowing is assumed. The observational uncertainties of the cosmographic parameters are propagated simultaneously: in each Monte Carlo realization, $q_0$ and $j_0$ are drawn from asymmetric Gaussian distributions whose left and right standard deviations reproduce the $1\sigma$ errors quoted in eqs.~\eqref{q0_values} and~\eqref{j0_observational}, and $\Delta$ is drawn from the uniform prior. Since the adopted $q_0$ and $j_0$ come from independent analyses \cite{Mukherjee:2020vkx,AlMamon:2018uby}, they are treated as uncorrelated; imposing a hypothetical correlation $|\rho| \le 0.5$ between them changes the resulting $\sigma_\delta$ by less than $11\%$. Each realization yields a point $(\Delta_i, \delta_i)$ with $\delta_i = [4-\alpha(q_{0,i}, j_{0,i})]/(2+\Delta_i)$; the small fraction of realizations violating the physical requirement $\delta > 0$ ($4.8\%$ for the reference set, originating from the far tails where $q_0$ approaches the de Sitter pole of eq.~\eqref{alpha_qj}) is discarded. Finally, from the resulting distribution of the parameter $\delta$, we derive an estimate in the form $\delta_0 \pm \sigma_\delta$, where $\sigma_\delta$ is half the width of the 68\% interval.

\begin{figure}[tbp]
	\centering
	\includegraphics[width=\textwidth]{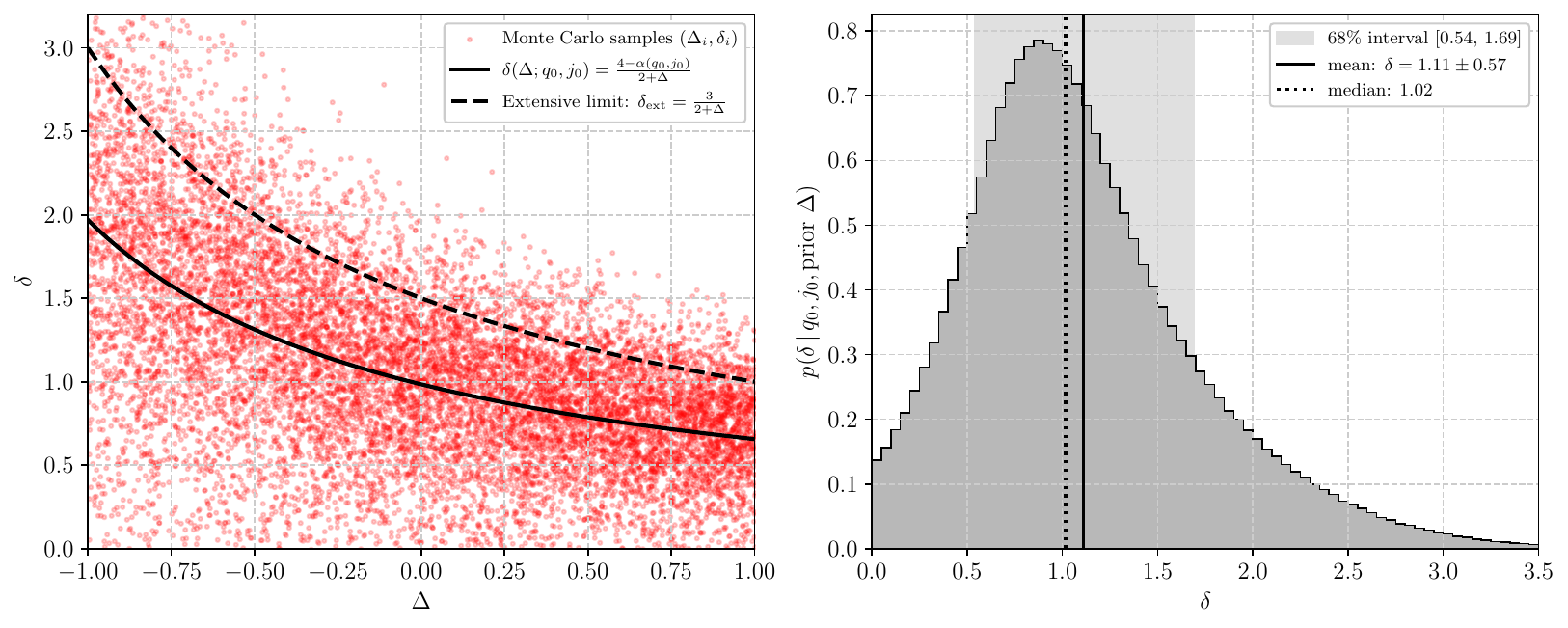}
	\caption{Left panel: relation $\delta(\Delta)$ for the reference cosmographic parameters $q_0 = -0.580^{+0.055}_{-0.063}$ and $j_0 = 0.742^{+0.106}_{-0.094}$. The solid curve shows the Barrow--Tsallis model at the central values, and the dashed curve shows the extensive limit. The red points show the joint Monte Carlo sample, in which $q_0$, $j_0$ and $\Delta$ are varied simultaneously; the spread of the points around the central curve reflects the observational uncertainties of the cosmographic parameters. Right panel: the resulting distribution of the nonextensivity parameter $\delta$; the solid and dotted vertical lines mark the mean and the median, and the shaded band shows the 68\% interval, eq.~\eqref{delta_estimate}.}
	\label{fig:MC_delta}
\end{figure}

For the reference set, $q_0 = -0.580^{+0.055}_{-0.063}$ and $j_0 = 0.742^{+0.106}_{-0.094}$, the joint Monte Carlo propagation yields
\begin{equation}\label{delta_estimate}
	\delta = 1.11 \pm 0.57
\end{equation}
(median $1.02$, 68\% interval $[0.54, 1.69]$); see figure~\ref{fig:MC_delta}. It is instructive to decompose the error budget by varying one source of uncertainty at a time: the observational error of the jerk parameter alone gives $\sigma_\delta \simeq 0.39$, the uniform prior on $\Delta$ over the full interval $[-1,1]$ gives $\sigma_\delta \simeq 0.38$, and the error of the deceleration parameter gives $\sigma_\delta \simeq 0.24$; the jerk parameter thus provides the largest observational contribution, in accordance with the discussion above. Note also that the prior-independent combination fixed by cosmography alone is $\delta(2+\Delta) = 4 - \alpha(q_0,j_0) = 1.97 \pm 0.95$. Table~\ref{tab:MC_delta} collects the estimates for all four determinations of $q_0$ in eq.~\eqref{q0_values} and for both the full prior range $\Delta \in [-1,1]$ and the narrower range $\Delta \in [-0.5,0.5]$ considered for comparison: the central value of $\delta$ is stable at the level of $\lesssim 15\%$ across the four datasets --- well within the quoted uncertainties --- and the choice of the prior range affects mainly the width of the distribution. This result indicates a moderate nonextensivity of the Barrow--Tsallis entropy, which is statistically consistent with the Bekenstein value $\delta = 1$ (equivalently, with the standard scaling $\alpha = 2$ of the holographic density). We note that our estimate is in good agreement with the values reported in a recent study \cite{Yarahmadi:2025luc}, which also employs statistical methods based on machine learning approaches to determine the nonextensivity parameter.

\begin{table}[tbp]
	\centering
	\begin{tabular}{lccc}
		\hline
		Set & $q_0$ \cite{Mukherjee:2020vkx} & $\delta$, $\Delta \in [-1,1]$ & $\delta$, $\Delta \in [-0.5,0.5]$ \\
		\hline
		N1 (CCB+SN) & $-0.573^{+0.041}_{-0.042}$ & $1.15 \pm 0.56$ & $1.06 \pm 0.44$ \\
		N2 (CCM+SN) & $-0.580^{+0.055}_{-0.063}$ & $1.11 \pm 0.57$ & $1.03 \pm 0.47$ \\
		N3 (CCB+SN+BAO) & $-0.533^{+0.038}_{-0.038}$ & $1.28 \pm 0.55$ & $1.19 \pm 0.40$ \\
		N4 (CCM+SN+BAO) & $-0.574^{+0.044}_{-0.045}$ & $1.14 \pm 0.56$ & $1.06 \pm 0.44$ \\
		\hline
	\end{tabular}
	\caption{Monte Carlo estimates of the nonextensivity parameter $\delta$ (mean $\pm$ half of the 68\% interval) for the four determinations of $q_0$ from ref.~\cite{Mukherjee:2020vkx} [eq.~\eqref{q0_values}], with $j_0 = 0.742^{+0.106}_{-0.094}$ \cite{AlMamon:2018uby} in all cases. The third column corresponds to the uniform prior over the full physically allowed range of the Barrow parameter, and the fourth column to the narrower range used for comparison. The labels N1--N4 follow ref.~\cite{Mukherjee:2020vkx}: cosmic chronometers with the BC03 (CCB) or MaStro (CCM) stellar-population-synthesis calibration, combined with the Pantheon supernovae, without or with BAO data.}
	\label{tab:MC_delta}
\end{table}

\section{Fractional holographic dark energy}

We also applied our cosmographic reconstruction method to evaluate the feasibility of using fractional derivatives to describe the late evolution of the Universe. Following \cite{Trivedi:2024inb}, we analyzed a scenario based on fractional holographic dark energy (FHDE). The fractional derivative generalizes the usual integer order of differentiation to rational, real, or complex numbers.

Using the IR--UV correspondence, one can construct the density of 
fractional holographic dark energy:
\begin{equation}
	\rho_{de} = 3c_{\mathrm{fr}}^{2}\,L^{\frac{2-3\eta}{\eta}},
	\end{equation}
where $\eta$ is a fractional exponent bounded by the interval $1 \le \eta \le 2$ and $c_{\mathrm{fr}}$ is the dimensionless normalization parameter of the FHDE model. The subscript ``fr'' only distinguishes the fractional model; $c_{\mathrm{fr}}$ denotes the same normalization constant conventionally written as $c$ in the standard holographic limit $\eta \to 2$.

For $\eta = 2$, the usual density of holographic (entropic) dark energy 
is restored. Choosing the infrared cutoff in the form $L = H^{-1}$, we obtain
\begin{equation}
	\rho_{de} = 3c_{\mathrm{fr}}^{2} H^{\frac{3\eta-2}{\eta}}
\end{equation}

The functional form of fractional holographic dark energy 
$\rho_{de} \propto H^\sigma, \sigma \equiv (3\eta - 2)/\eta$ allows one 
to use the obtained relation for the holographic Barrow dark energy, which 
has a similar dependence on the Hubble parameter, to determine the parameter
\begin{equation}
	\sigma \equiv \frac{3\eta - 2}{\eta} = \frac{3(1 - j)}{(1 + q)^2 (1 - 2q)}
\end{equation}
For any given value of $q$, the allowable values of $j$ form an interval 
$j_{\min}(q) \le j \le j_{\max}(q)$. It is convenient to introduce the notation 
\begin{equation}
	j_{*}(q) = \frac{1}{3}(1 + q)^2 (1 - 2q)
\end{equation}

The boundary values are defined as:
\begin{equation}
	j_{\min}(q) = \min\{1 - 2j_{*}(q), 1 - j_{*}(q)\}, \quad j_{\max}(q) = \max\{1 - 2j_{*}(q), 1 - j_{*}(q)\}
\end{equation}
In the late-time Universe ($q \lesssim 1/2$), these expressions simplify to:
\begin{equation}
	j_{\min}(q) = 1 - 2j_{*}(q), \quad j_{\max}(q) = 1 - j_{*}(q)
\end{equation}
where the function $j_{*}(q)$ precisely defines the width of the interval: $j_{*}(q) = j_{\max} - j_{\min}$.

\begin{figure}[tbp]
	\centering
	\includegraphics[width=0.8\textwidth]{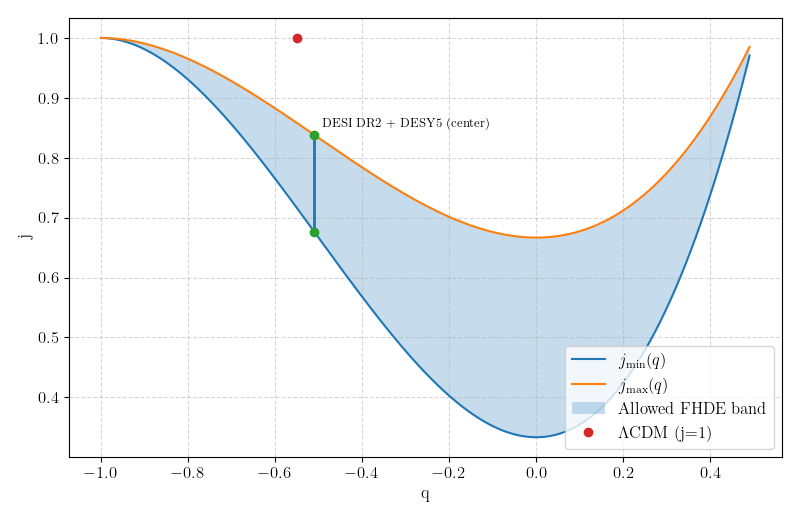}
	\caption{The allowable region for the jerk parameter $j$ as a function of the deceleration parameter $q$ in the fractional holographic dark energy model. The shaded area represents the parameter space where the fractional exponent $\eta$ lies within the physical interval $[1, 2]$. The markers show the $\Lambda$CDM point and the DESI DR2 + DES Y5 estimate \cite{DESI2025DR2I,DESI2025DR2II,DES2024SN5YRcosmo,DES2024SN5YRlightcurves}.}
	\label{fig:MC_delta_FHDE}
\end{figure}

This relation illustrates the vanishing of the interval ($j_{*}(q) \to 0$) in the de Sitter limit ($q \to -1$), the expansion of the allowed region as the system evolves towards $q \to 0$, and the subsequent contraction of the interval as $q \to 1/2$ (see figure~\ref{fig:MC_delta_FHDE}).

In \cite{Trivedi:2024inb}, an expression of the form $\rho_{\mathrm{de}}=3c^{2}L^{(2-3\eta)/\eta}$ was derived for the density of FHDE (where we have identified their fractional parameter with our $\eta$ and their normalization constant $c$ with our $c_{\mathrm{fr}}$). For the choice of the Hubble cutoff, $L=H^{-1}$, this expression reduces to the power law $\rho \propto H^{(3\eta-2)/\eta}$. This particular case corresponds directly to our parameterization $\rho_{\mathrm{de}}^{(\mathrm{fr})}=3c_{\mathrm{fr}}^{2}H^{\sigma}$, where $\sigma=(3\eta-2)/\eta$. Thus, the FHDE model with $L=H^{-1}$, considered therein, naturally fits into the general class of power-law models $\rho\propto H^{\alpha}$, which we employ as a unifying framework.

While ref.~\cite{Bidlan:2026xaa} primarily focuses on the classification of future evolution scenarios (Big Rip, Little Rip, and Pseudo Rip) for given types of infrared cutoff scales and interaction forms in the dark sector, the key aspect of the present work is to establish a connection between the fractional FHDE parameter and the cosmographic quantities $(q,j)$. The relation derived here, $\sigma=\sigma(q,j)$, combined with the allowable interval $j\in[j_{\min}(q),j_{\max}(q)]$, provides a tool for confronting the $\alpha$ values adopted in that work with observational constraints. This approach allows for a substantial refinement of the physically viable region of the model's parameter space.

It is important to note that the observed narrowing of the allowed interval for the jerk parameter, $j_{*}(q)\to 0$, in the limit $q\to -1$ (the de Sitter limit) is in good agreement with the interpretation of the pseudo-rip regime in ref.~\cite{Bidlan:2026xaa} as an asymptotic approach toward de Sitter evolution. However, it should be emphasized that their conclusions for the case $L=H^{-1}$ depend crucially on the postulated ansatz for $H(t)$ and the specific form of the interaction function $Q$. In this context, our cosmographic approach allows us to formulate constraints on $\sigma$ and diagnostic criteria in terms of $(q,j)$ without specifying a global ansatz for the expansion history, but within the assumed $L=H^{-1}$ power-law FHDE class. This provides a direct verification of the kinematic compatibility of the considered ``rip-like'' regimes with the dynamics of the late-time Universe.

\section{IR--UV correspondence and microscopic length scale}
The preceding cosmographic reconstruction yields a direct constraint on the Barrow--Tsallis entropy parameters. The Barrow parameter $\Delta$ encodes the microscopic deformation of the horizon geometry, whereas the Tsallis parameter $\delta$ characterizes the nonadditivity of the horizon entropy associated with the long-range nature of gravity. Within the spatially flat, non-interacting dust plus power-law holographic-dark-energy model with constant $\alpha$ and $\beta$, the central result is the exact algebraic relation
\begin{equation}\label{BT_cosmographic_constraint}
	\left( 1 + \frac{\Delta}{2} \right) \delta = 2 - \frac{1}{2}\alpha(q_0, j_0).
\end{equation}

The right-hand side depends only on the current kinematic cosmographic parameters $q = -\frac{\ddot{a}}{aH^2}$ and $j = \frac{\dddot{a}}{aH^3}$. No approximation enters the algebraic inversion itself. However, the relation is conditional on the adopted dark-energy ansatz and conservation laws; within this fixed model class, its observational uncertainty is set by the uncertainties of $q_0$ and $j_0$. The IR--UV correspondence used below is an additional physical hypothesis and is not part of the algebraic exactness \cite{Bolotin:2023wiw}. Let us examine this hypothesis in more detail, initially using the Bekenstein--Hawking entropy as an example for simplicity. 

In any effective quantum field theory defined in a spatial domain of 
characteristic size $L$ and using the ultraviolet cutoff $\Lambda$, the 
entropy is $S \propto L^3 \Lambda^3$. If, using the IR--UV correspondence, 
we assume that the entropy of any object of linear size $L$ in such a 
theory must be less than the entropy of a black hole of the same size, 
then \cite{Ng:2003jk,Christiansen:2009bz,YuLBolotin2020PhysicsOL,Bolotin:2024gjd}

\begin{equation}\label{S_le_S_BH}
	S \le S_{BH} \approx \left( \frac{L}{l_p} \right)^2
\end{equation}
Inequality \eqref{S_le_S_BH} can be transformed into a constraint on the inverse value of the UV cutoff:
\begin{equation}
	\Lambda^{-1} \ge \left( L l_p^2 \right)^{1/3}
\end{equation}
explicitly linking the macro ($L$) and micro ($\Lambda^{-1}$) scales. If we 
identify the inverse value of the UV cutoff $\Lambda^{-1}$ with the maximum 
limit of the accuracy of length measurement $\delta L$, then the last 
inequality is transformed into a fundamental inequality \cite{Karolyhazy:1966zz}:
\begin{equation}\label{delta_L}
	\delta L \ge \left( L l_p^2 \right)^{1/3}
\end{equation}

Carrying out a similar procedure for the Barrow--Tsallis entropy, we find:
\begin{equation}\label{delta_LL}
\Lambda^{-1} \ge \gamma^{-1/3} L^{1-\frac{\delta}{3}(2+\Delta)}  l_{p}^{\frac{\delta}{3}(2+\Delta)}
\end{equation}

Using \eqref{BT_cosmographic_constraint}, eq.~\eqref{delta_LL} can be rewritten in terms of the cosmographically reconstructed exponent $\alpha(q_0,j_0)$:
\begin{equation}\label{delta_LL_cosmo}
	\Lambda^{-1} \ge \gamma^{-1/3}
	L^{\frac{\alpha(q_0,j_0)-1}{3}}
	l_p^{\frac{4-\alpha(q_0,j_0)}{3}} .
\end{equation}
Identifying $\Lambda^{-1}$ with the limiting uncertainty of length measurement yields
\begin{equation}\label{delta_L_BT}
	\delta L_{\mathrm{BT}} \ge \gamma^{-1/3}
	L^{\frac{\alpha(q_0,j_0)-1}{3}}
	l_p^{\frac{4-\alpha(q_0,j_0)}{3}} .
\end{equation}
For $\Delta=0$ and $\delta=1$, one has $\alpha=2$. With the scaling convention $\gamma=1$ adopted above, eq.~\eqref{delta_L_BT} reduces to eq.~\eqref{delta_L}; for arbitrary $\gamma$, the standard exponents are unchanged and only the overall prefactor is multiplied by $\gamma^{-1/3}$. The Barrow--Tsallis deformation alters the scaling exponents of both the macroscopic scale $L$ and the Planck scale $l_p$ in the microscopic length-uncertainty bound. Consequently, the cosmographic constraint on the macroscopic nonextensivity parameter determines the quantum-foam scaling law within the IR--UV correspondence.

To provide an explicit numerical example, we substitute the reference values $q_0 = -0.580$ and $j_0 = 0.742$ adopted in section~1 into eq.~\eqref{alpha_q0j0}, obtaining $\alpha(q_0, j_0) \simeq 2.031$ (the joint propagation of the observational errors gives $\alpha = 2.12 \pm 0.95$). Using this value in eq.~\eqref{delta_L_BT}, with the scaling convention $\gamma=1$ adopted above, yields
\begin{equation}\label{delta_L_BT_num}
	\delta L_{\mathrm{BT}} \gtrsim L^{0.344}\, l_p^{0.656},
\end{equation}
to be compared with the Karolyhazy exponents $(1/3,\, 2/3)$. Taking for the macroscopic scale the present Hubble radius, $L = c H_0^{-1} \simeq 1.32 \times 10^{26}$~m (for $H_0 = 70$~km\,s$^{-1}$\,Mpc$^{-1}$), the standard bound \eqref{delta_L} gives $\delta L \gtrsim (L l_p^2)^{1/3} \simeq 3.3 \times 10^{-15}$~m, whereas the Barrow--Tsallis bound \eqref{delta_L_BT_num} gives $\delta L_{\mathrm{BT}} \gtrsim 1.4 \times 10^{-14}$~m. The ratio of the two bounds is controlled by a single exponent,
\begin{equation}\label{dL_ratio}
	\frac{\delta L_{\mathrm{BT}}}{\delta L} = \left( \frac{L}{l_p} \right)^{\frac{\alpha(q_0, j_0) - 2}{3}},
\end{equation}
and since $L/l_p \sim 10^{61}$, even the modest displacement $\alpha - 2 \simeq 0.03$ implied by the current central values rescales the quantum-foam length by a factor of a few, while the present observational uncertainty $\sigma_\alpha \simeq 0.95$ (dominated by the jerk parameter) leaves the bound undetermined to within many orders of magnitude. Conversely, eq.~\eqref{dL_ratio} shows that determining the microscopic scale to within a factor of $e$ requires $\sigma_\alpha \lesssim 3/\ln(L/l_p) \simeq 0.02$, i.e.\ an improvement of the accuracy of $\alpha(q_0, j_0)$ --- and hence, primarily, of the jerk parameter --- by a factor of $\sim 45$, precisely the kind of progress anticipated from next-generation gravitational-wave cosmography \cite{Chen:2024gdn}.

\section{Conclusions}
The viability of any cosmological model is determined by its ability to 
reproduce observational results. A necessary preliminary step preceding this 
procedure is the determination of the model's parameters. In the present work, 
this step is implemented as an inverse cosmographic reconstruction: the 
parameters are fixed by the local kinematic data. We emphasize that this reconstruction is algebraically exact within the adopted model class --- the spatially flat, non-interacting dust plus power-law holographic dark energy with constant parameters --- and, in the microscopic-length analysis, relies on the IR--UV correspondence as an additional physical hypothesis; the relations obtained here are thus exact in this conditional sense rather than model-independent statements. The density of 
holographic dark energy $\rho_{de} \propto H^\alpha$ constructed using the 
Barrow--Tsallis entropy is fixed by a combination of original entropy 
parameters $\alpha = 4 - 2\delta - \delta \Delta$. 
We expressed this combination in terms of the current values of the observed 
cosmographic parameters $q_0, j_0$. This relationship transforms the 
two-parameter Barrow--Tsallis entropy model of dark energy with parameters 
$\Delta$ and $\delta$ into a one-parameter model $[\Delta, \delta(\Delta)]$. 
This result, connecting the microscopic Barrow deformation ($\Delta$) and 
the macroscopic Tsallis nonextensivity ($\delta$), is an explicit realization 
of the IR--UV correspondence. In particular, after identifying $\Lambda^{-1}$ 
with the limiting length uncertainty, the Barrow--Tsallis constraint changes 
the scaling to 
$\delta L \gtrsim \gamma^{-1/3} L^{[\alpha(q_0,j_0)-1]/3} l_p^{[4-\alpha(q_0,j_0)]/3}$, 
which reduces to the standard Karolyhazy relation for $\Delta=0$, $\delta=1$, and the scaling convention $\gamma=1$. 
The joint Monte Carlo propagation of the observational errors of $(q_0, j_0)$, combined with marginalization over the uniform prior on $\Delta$ within the adopted interval $[-1,1]$, yields $\delta = 1.11 \pm 0.57$, statistically consistent with the Bekenstein value $\delta = 1$, with the jerk parameter giving the largest observational contribution to the error budget. Our estimate of the permissible ranges of the model parameters is consistent 
with the estimate of the nonextensivity parameter obtained using machine 
learning. We also applied our cosmographic reconstruction method to evaluate the feasibility of using fractional 
derivatives to describe the late evolution of the Universe.

\bibliographystyle{JHEP}
\bibliography{bibli2}

\end{document}